\documentclass{article}
\usepackage{amsfonts}
\usepackage{amsmath}

\input{tcilatex}
\begin{document}

\title{Comment on "the quantum dynamics for general time-dependent three
coupled oscillators"}
\author{\textbf{\ Zerimeche Rahma }$^{a\text{,}b,1}$\textbf{, Mana Naima}$%
^{a,2}$ \and \textbf{\ and \ Maamache Mustapha}$^{a,3}$ {\small \thanks{$%
^{1} $zerimecherahma@gmail.com,$^{2}$na3ima\_mn@hotmail.fr,$^{3}$%
maamache@univ-setif.dz }} \\
$^{a}${\small Laboratoire de Physique Quantique et Syst\`{e}mes Dynamiques,}%
\\
{\small Facult\'{e} des Sciences, Ferhat Abbas S\'{e}tif 1, S\'{e}tif 19000,
Algeria.}\\
$^{b}${\small Physics Department, University of Jijel, BP 98, Ouled Aissa, }%
\\
{\small 18000 Jijel, Algeria.}}
\date{}
\maketitle

\begin{abstract}
In a recent paper, Hassoul et al.\cite{Hassoul2022} , the authors proposed
an analysis of the quantum dynamics for general time-dependent three coupled
oscillators \ through an approach based on their decouplement using the
unitary transformation method. Thus, to diagonalize the transformed
Hamiltonian, they introduce a new unitary operator corresponding to a
three-dimensional rotation parameterized by Euler angles. Through this
procedure, Hassoul et al. \cite{Hassoul2022} claim that the coupled
oscillatory subsystems are completely decoupled. This last approach is
partly wrong. In this brief note, we show that their method is indeed not
correct and we try to explain what\ truly lies behind their mistakes. We
also propose a brief discussion on an alternative method that might achieve
satisfactory results.
\end{abstract}

\section{Introduction}

The study of the dynamic behavior of oscillating systems is a central issue
in applied sciences and mathematics. The harmonic oscillator model has been
studied so far for a wide range of mechanical systems. However, if the
oscillator interacts with the environment or with other oscillators, the
associated systems cannot be isolated.

Coupled oscillators are connected in such a way that energy can be
transferred between them, and their motion is usually complex. Therefore,
one can find a coordinate frame in which each oscillator oscillates with a
well-defined frequency. Knowing that the coupled oscillators are very
complex from the dynamic point of view and in particular when their
parameters depend on time and/or when the number of coupled oscillatory
subsystems is greater than two. Despite the key role of coupled oscillatory
models in general mechanical descriptions, their studies have been mainly
performed for cases where oscillator parameters, such as masses and
frequencies, are independent of time. Given the need to develop the dynamics
of non-conservative oscillatory physical systems, Hassoul et al \cite%
{Hassoul2022} consider a system of three coupled oscillators where the
parameters of the Hamiltonian are arbitrary time-dependent functions, their
main idea is that the Hamiltonian can be written in a diagonal form and that
the exact solutions of the Schr\"{o}dinger equation can be obtained in a
simple way. In this brief note, we show that the method used in Ref. \cite%
{Hassoul2022} is not adapted to the system studied because the rotation
matrix as a function of Euler angles cannot diagonalize the Hamiltonian (13
) Ref. \cite{Hassoul2022}. We end with a brief discussion of an alternative
method that gives satisfactory results for diagonalizing the Hamiltonian.

\section{A short review: Rotation matrix and diagonalization of Hamiltonian}

We begin this section by showing, with a direct computation, that the paper 
\cite{Hassoul2022} contains an essential (and, in fact, trivial) mistake,
which makes their results incorrect. The considered system is described by
the time-dependent Hamiltonian (Eq. (13)) of Ref. \cite{Hassoul2022}:%
\begin{eqnarray}
H(t) &=&\frac{1}{2}\overset{3}{\underset{i=1}{\sum }}\left[ P_{i}^{2}+\varpi
_{i}^{2}(t)X_{i}^{2}\right]
+K_{12}X_{1}X_{2}+K_{13}X_{1}X_{3}+K_{23}X_{2}X_{3}  \notag \\
&=&\frac{1}{2}\overset{3}{\underset{i,j=1}{\sum }}P_{i}\delta _{ij}P^{j}+%
\frac{1}{2}\overset{3}{\underset{i,j=1}{\sum }}X_{i}\Gamma _{ij}X^{j},
\label{1}
\end{eqnarray}%
where%
\begin{equation}
\Gamma (t)=\left( 
\begin{array}{ccc}
\varpi _{1}^{2}(t) & K_{12} & K_{13} \\ 
K_{12} & \varpi _{2}^{2}(t) & K_{23} \\ 
K_{13} & K_{23} & \varpi _{3}^{2}(t)%
\end{array}%
\right) ,  \label{1'}
\end{equation}%
and 
\begin{equation}
\varpi _{i}^{2}(t)=\frac{1}{4}\left( \frac{\dot{m}_{i}^{2}(t)}{m_{i}^{2}(t)}-%
\frac{2\ddot{m}_{i}(t)}{m_{i}(t)}\right) +\frac{c_{i}(t)}{m_{i}\left(
t\right) },  \label{2}
\end{equation}%
\begin{equation}
K_{12}=\frac{c_{12}(t)}{2\sqrt{m_{1}(t)m_{2}(t)}},\ K_{13}=\frac{c_{13}(t)}{2%
\sqrt{m_{1}(t)m_{3}(t)}}\ \text{and\ }K_{23}=\frac{c_{23}(t)}{2\sqrt{%
m_{2}(t)m_{3}(t)}},  \label{2'}
\end{equation}%
\ \ and the parameters $m_{i}\left( t\right) $ $\left( i=1,2,3\right) $ and $%
c_{i}(t)$ $\left( i=1,2,3\right) $ are arbitrary functions of time.

The eigenvalues of the $3\times 3$ matrix $\Gamma (t)$ are 
\begin{eqnarray}
\Omega _{1}^{2} &=&\frac{1}{3}\left[ \left( \mathcal{\varpi }_{1}^{2}+%
\mathcal{\varpi }_{2}^{2}+\mathcal{\varpi }_{3}^{2}\right) +2\Omega \cos
\left( \frac{\Phi }{3}\right) \right] ,  \notag \\
\Omega _{2}^{2} &=&\frac{1}{3}\left[ \left( \mathcal{\varpi }_{1}^{2}+%
\mathcal{\varpi }_{2}^{2}+\mathcal{\varpi }_{3}^{2}\right) +2\Omega \cos
\left( \frac{\Phi +2\pi }{3}\right) \right] ,  \notag \\
\Omega _{3}^{2} &=&\frac{1}{3}\left[ \left( \mathcal{\varpi }_{1}^{2}+%
\mathcal{\varpi }_{2}^{2}+\mathcal{\varpi }_{3}^{2}\right) +2\Omega \cos
\left( \frac{\Phi -2\pi }{3}\right) \right] ,  \label{3}
\end{eqnarray}

\begin{equation}
\Phi =\arccos \left( \frac{\Delta }{2\sqrt{\Omega ^{3}}}\right) ,  \label{3'}
\end{equation}%
the expressions of $\Omega $ and $\Delta $ are 
\begin{equation}
\Omega =\frac{1}{2}\left[ \left( \mathcal{\varpi }_{1}^{2}-\mathcal{\varpi }%
_{2}^{2}\right) ^{2}+\left( \mathcal{\varpi }_{1}^{2}-\mathcal{\varpi }%
_{3}^{2}\right) ^{2}+\left( \mathcal{\varpi }_{2}^{2}-\mathcal{\varpi }%
_{3}^{2}\right) ^{2}\right] +3\left( K_{23}^{2}+K_{13}^{2}+K_{12}^{2}\right)
,  \label{4}
\end{equation}%
and%
\begin{eqnarray}
\Delta &=&18\left( \mathcal{\varpi }_{1}^{2}\mathcal{\varpi }_{2}^{2}%
\mathcal{\varpi }_{3}^{2}+3K_{12}K_{13}K_{23}\right) +2\left( \mathcal{%
\varpi }_{1}^{3}+\mathcal{\varpi }_{3}^{3}+\mathcal{\varpi }_{2}^{3}\right)
+9\left( \mathcal{\varpi }_{1}^{2}+\mathcal{\varpi }_{2}^{2}+\mathcal{\varpi 
}_{3}^{2}\right) \left( K_{23}^{2}+K_{13}^{2}+K_{12}^{2}\right)  \notag \\
&&-3\left( \mathcal{\varpi }_{1}^{2}+\mathcal{\varpi }_{2}^{2}\right) \left( 
\mathcal{\varpi }_{1}^{2}+\mathcal{\varpi }_{3}^{2}\right) \left( \mathcal{%
\varpi }_{2}^{2}+\mathcal{\varpi }_{3}^{2}\right) -27\left( \mathcal{\varpi }%
_{1}^{2}K_{23}^{2}+\mathcal{\varpi }_{2}^{2}K_{13}^{2}+\mathcal{\varpi }%
_{3}^{2}K_{12}^{2}\right) .  \label{4'}
\end{eqnarray}

Knowing, from classical mechanics, that an arbitrary rotation of a rigid
body can be expressed in terms of three consecutive rotations, called the
Euler rotations. In order to diagonalize the matrix Hamiltonian $H(t)$ (\ref%
{1}) Hassoul et al perform a unitary transformation\ that corresponds to a
three-dimensional rotation parameterized by three Euler angles $(\phi
,\theta ,\varphi )$ introduced as%
\begin{eqnarray}
\mathbb{R}
&=&%
\mathbb{R}
_{X_{1}}(\phi )%
\mathbb{R}
_{X_{2}}(\theta )%
\mathbb{R}
_{X_{3}}(\varphi ),  \notag \\
&=&\left( 
\begin{array}{ccc}
1 & 0 & 0 \\ 
0 & \cos \phi & -\sin \phi \\ 
0 & \sin \phi & \cos \phi%
\end{array}%
\right) \left( 
\begin{array}{ccc}
\cos \theta & 0 & \sin \theta \\ 
0 & 1 & 0 \\ 
-\sin \theta & 0 & \cos \theta%
\end{array}%
\right) \left( 
\begin{array}{ccc}
\cos \varphi & -\sin \varphi & 0 \\ 
\sin \varphi & \cos \varphi & 0 \\ 
0 & 0 & 1%
\end{array}%
\right) ,  \notag \\
&=&\left( 
\begin{array}{ccc}
\cos \theta \cos \varphi & -\cos \theta \sin \varphi & \sin \theta \\ 
\sin \phi \sin \theta \cos \varphi +\cos \phi \sin \varphi & \cos \phi \cos
\varphi -\sin \phi \sin \varphi \sin \theta & -\sin \phi \cos \theta \\ 
-\cos \theta \sin \theta \cos \varphi +\sin \varphi \sin \phi & \sin \phi
\cos \varphi +\sin \varphi \cos \phi \sin \theta & \cos \phi \cos \theta%
\end{array}%
\right) ,  \label{5}
\end{eqnarray}%
seeing the rotation matrix $%
\mathbb{R}
$, we can determine how vectors transform under rotations; in a
three-dimensional space, vectors $X$ and $P$ are rotated as $q$ = $%
\mathbb{R}
^{-1}X$ and $p$ = $%
\mathbb{R}
^{-1}P$. It is therefore more convenient, in quantum mechanics, to
parametrize, as in classical mechanics in terms of the three Euler angles $%
(\phi ,\theta ,\varphi )$, the rotation operator $\Lambda (t)$ that
corresponds to the rotation matrix $%
\mathbb{R}
_{X_{1}}(\phi )%
\mathbb{R}
_{X_{2}}(\theta )%
\mathbb{R}
_{X_{3}}(\varphi )$ in the form%
\begin{equation}
\Lambda (t)=\exp \left[ i\phi (t)J_{1}\right] \exp \left[ i\theta (t)J_{2}%
\right] \exp \left[ i\varphi (t)J_{3}\right] .  \label{6}
\end{equation}

The components of rotation generators obey the commutation relations%
\begin{equation}
\left[ J_{i},J_{j}\right] =ih\varepsilon _{ijk}J_{k},\text{ \ \ }\left[
J_{i},X_{j}\right] =ih\varepsilon _{ijk}X_{k},\text{ \ \ \ }\left[
J_{i},P_{j}\right] =ih\varepsilon _{ijk}P_{k},\text{\ }  \label{6'}
\end{equation}%
$\varepsilon _{ijk}$ is the Levi-Civita symbol.

In order to eliminate the coupled terms $X_{i}X_{j}$, Hassoul et al \cite%
{Hassoul2022} evaluate $%
\mathbb{R}
^{-1}(t)\Gamma (t)%
\mathbb{R}
(t)$ and affirm that it is possible to verify the relation $%
\mathbb{R}
(t)\left\{ \text{diag}\left[ \Omega _{1}^{2},\Omega _{2}^{2},\Omega _{3}^{2}%
\right] \right\} 
\mathbb{R}
^{-1}(t)$ \ in terms of the new diagonal matrix%
\begin{equation}
\Gamma (t)=%
\mathbb{R}
(t)\left( 
\begin{array}{ccc}
\Omega _{1}^{2} & 0 & 0 \\ 
0 & \Omega _{2}^{2} & 0 \\ 
0 & 0 & \Omega _{3}^{2}%
\end{array}%
\right) 
\mathbb{R}
^{-1}(t).  \label{matK}
\end{equation}

This statement is incorrect and to be convinced, it suffices to evaluate $%
\mathbb{R}
^{-1}(t)\Gamma (t)%
\mathbb{R}
(t)$. So let's embark on the calculation of%
\begin{equation}
\mathbb{R}
^{-1}(t)\Gamma (t)%
\mathbb{R}
(t)=\left( 
\begin{array}{ccc}
M_{11} & M_{12} & M_{13} \\ 
M_{21} & M_{22} & M_{23} \\ 
M_{31} & M_{32} & M_{33}%
\end{array}%
\right) ,  \label{7'}
\end{equation}%
where the coefficients $M_{ij}$ are given in the appendix. To get the
diagonal matrix diag $\left[ \Omega _{1}^{2},\Omega _{2}^{2},\Omega _{3}^{2}%
\right] $, we put $M_{ii}=\Omega _{i}^{2}$ and $M_{ij}=0$ for $\left( i\neq
j\right) $ and in this case we deduce that $K_{12}=K_{13}=K_{23}=0$ which
implies that $\varpi _{1}^{2}=\varpi _{2}^{2}=\varpi _{3}^{2}$ and $\Omega
_{1}^{2}=\Omega _{2}^{2}=\Omega _{3}^{2}=\varpi _{1}^{2}(t)$ which does not
coincide with Eq. (36) of Ref.\cite{Hassoul2022}, therefore the quantum
system described by Hamiltonian (\ref{1}) has not been solved as claimed by
Hassoul et al \cite{Hassoul2022}. In order to remedy this situation, we will
adopt the approach of diagonalization\ described in \cite{Park} but before
that we shall emphasize that the rotated-coordinate column vector $X_{%
\mathbb{R}
}$ are%
\begin{equation}
\Lambda ^{-1}(t)X_{1}\Lambda (t)=\cos \theta \cos \varphi X_{1}+\cos \theta
\sin \varphi X_{2}-\sin \theta X_{3}\text{ ,}
\end{equation}%
\begin{equation}
\Lambda ^{-1}(t)X_{2}\Lambda (t)=\left( \sin \phi \sin \theta \cos \varphi
-\sin \varphi \cos \phi \right) X_{1}+\left( \sin \phi \sin \theta \sin
\varphi +\cos \varphi \cos \phi \right) X_{2}+\sin \phi \cos \theta X_{3}%
\text{ ,}
\end{equation}%
\begin{equation}
\Lambda ^{-1}(t)X_{3}\Lambda (t)=\left( \sin \theta \cos \phi \cos \varphi
+\sin \phi \sin \varphi \right) X_{1}+\left( \sin \theta \cos \phi \sin
\varphi -\sin \phi \cos \varphi \right) X_{2}+\cos \phi \cos \theta X_{3%
\text{ }}\text{ ,}
\end{equation}%
which we can write as%
\begin{equation}
\Lambda ^{-1}(t)X_{k}\Lambda (t)=\sum_{k=1}^{3}\bar{%
\mathbb{R}%
}_{ik}X_{k}\text{,}
\end{equation}%
where the matrix $\bar{%
\mathbb{R}%
}$ has the form%
\begin{equation}
\bar{%
\mathbb{R}%
}=\left( 
\begin{array}{ccc}
\cos \theta \cos \varphi & \cos \theta \sin \varphi & -\sin \theta \\ 
\sin \phi \sin \theta \cos \varphi -\sin \varphi \cos \phi & \sin \phi \sin
\theta \sin \varphi +\cos \varphi \cos \phi & \sin \phi \cos \theta \\ 
\sin \theta \cos \phi \cos \varphi +\sin \phi \sin \varphi & \sin \theta
\cos \phi \sin \varphi -\sin \phi \cos \varphi & \cos \phi \cos \theta%
\end{array}%
\right) ,
\end{equation}%
we note that this matrix is different from the matrix $%
\mathbb{R}
$ given in \cite{Hassoul2022}\ with equation (22), and consequently formulae
of the rotated-coordinates (Eqs.(23,33 and 34)) are incorrect.

At the end of this section, let us note that the Hassoul et al's \cite%
{Hassoul2022} paper is largely inspired from the incoherent results of \cite%
{hab}.

\section{Brief discussion: the right diagonalization of Hamiltonian}

The existence of a basis of eigenvectors makes possible to diagonalize the
Hamitonian in equation (\ref{1}), then it is necessary to seek the
corresponding eigenvectors as 
\begin{equation}
v(t)=\frac{1}{\sqrt{3}}\left( 
\begin{array}{c}
1 \\ 
1 \\ 
1%
\end{array}%
\right) ,\text{ \ \ }v_{\pm }(t)=A_{\pm }\left( 
\begin{array}{c}
K_{12}-K_{23}\mp z \\ 
K_{13}-K_{12}\pm z \\ 
K_{23}-K_{13}%
\end{array}%
\right) ,  \label{8}
\end{equation}%
where%
\begin{eqnarray}
A_{\pm }(t) &=&\frac{1}{\sqrt{2z\left[ 2z-\left(
J_{13}+J_{23}-2J_{12}\right) \right] }}\text{ ,}  \label{8'} \\
z &=&\sqrt{K_{12}^{2}+K_{13}^{2}+K_{23}^{2}-\left(
K_{12}K_{13}+K_{12}K_{23}+K_{13}K_{23}\right) \text{ }}\text{.\ }  \notag
\end{eqnarray}

The eigenvalues associated to the eigenvectors $v(t)$\ and\ $v_{\pm }(t)$\
are 
\begin{eqnarray}
\lambda  &=&\frac{1}{3}\left[ \varpi _{1}^{2}(t)+\varpi _{2}^{2}(t)+\varpi
_{3}^{2}(t)+2\left( K_{12}+K_{13}+K_{23}\right) \right]   \notag \\
\lambda _{+} &=&\frac{1}{3}\left[ \varpi _{1}^{2}(t)+\varpi
_{2}^{2}(t)+\varpi _{3}^{2}(t)-\left( K_{12}+K_{13}+K_{23}\right) \right] +z
\notag \\
\lambda _{-} &=&\frac{1}{3}\left[ \varpi _{1}^{2}(t)+\varpi
_{2}^{2}(t)+\varpi _{3}^{2}(t)-\left( K_{12}+K_{13}+K_{23}\right) \right] -z
\end{eqnarray}%
and do not coincide with the eigenvalues $\Omega _{i}^{2}$\ (\ref{3})
(expressions (41-43) in \cite{Hassoul2022}).

Thus, $\Gamma (t)$ can be diagonalized as $\Gamma (t)$ = $\mathcal{U}%
^{+}(t)\Gamma _{D}(t)(t)\mathcal{U}(t)$, where 
\begin{equation}
\mathcal{U}(t)=\left( 
\begin{array}{ccc}
1/\sqrt{3} & 1/\sqrt{3} & 1/\sqrt{3} \\ 
A_{+}\left( K_{12}-K_{23}-z\right) & A_{+}\left( K_{13}-K_{12}+z\right) & 
A_{+}\left( K_{23}-K_{13}\right) \\ 
A_{-}\left( K_{12}-K_{23}+z\right) & A_{-}\left( K_{13}-K_{12}-z\right) & 
A_{-}\left( K_{23}-K_{13}\right)%
\end{array}%
\right) .  \label{9}
\end{equation}

Now, we introduce new coordinates 
\begin{equation}
\left( 
\begin{array}{c}
\tilde{q}_{1} \\ 
\tilde{q}_{2} \\ 
\tilde{q}_{3}%
\end{array}%
\right) =\mathcal{U}(t)\left( 
\begin{array}{c}
X_{1} \\ 
X_{2} \\ 
X_{3}%
\end{array}%
\right) ,\text{ \ }\left( 
\begin{array}{c}
\tilde{p}_{1} \\ 
\tilde{p}_{2} \\ 
\tilde{p}_{3}%
\end{array}%
\right) =\mathcal{U}(t)\left( 
\begin{array}{c}
p_{1} \\ 
p_{2} \\ 
p_{3}%
\end{array}%
\right) ,\text{\ }  \label{9'}
\end{equation}%
in terms of the new coordinates the Hamiltonian (\ref{1}) can be
diagonalized in the form 
\begin{eqnarray}
H_{1}(t) &=&\frac{1}{2}\overset{3}{\underset{i,j=1}{\sum }}P_{i}\mathcal{U}%
_{ij}^{+}(t)_{i}\delta _{ij}\mathcal{U}_{ji}(t)P^{j}+\frac{1}{2}\overset{3}{%
\underset{i,j=1}{\sum }}X_{i}\mathcal{U}_{ij}^{+}(t)\mathcal{U}%
_{ji}(t)\Gamma _{ij}\mathcal{U}_{ij}^{+}(t)\mathcal{U}_{ji}(t)X^{j}  \notag
\\
&=&\frac{1}{2}\overset{3}{\underset{i,=1}{\sum }}\tilde{p}_{i}^{2}+\frac{1}{2%
}\overset{3}{\underset{i,j=1}{\sum }\Omega _{i}^{2}}\tilde{q}_{i}^{2}\text{
\ }  \label{10}
\end{eqnarray}%
where $\tilde{p}_{i}$ are the conjugate momenta of $\tilde{q}_{i}.$

Finally let us note that the solutions to the original problem can not be
obtained\ because the author has ignored in the calculations the term which
comes from the time derivative of the unitary operator transformation (2.8)
in \cite{Park}.\ We believe that this apparently natural procedure is ill
founded. To transform back to the original variables $X_{i}$, we first note
that since $\mathcal{U}(t)$ performs the scale change (\ref{9'}), the states
are related by $\left\langle \mathbf{\tilde{q}}\right\vert =\left\langle 
\mathbf{X}\right\vert \mathcal{U}^{-1}(t)$ \cite{Br, Ji, M1, M2}.

{\LARGE Appendix: Derivation of the coefficients }$M_{ij}$%
\begin{eqnarray}
M_{11} &=&\cos ^{2}\theta \cos ^{2}\varphi \mathcal{\varpi }_{1}^{2}+\left(
\cos \phi \sin \varphi +\sin \theta \sin \phi \cos \varphi \right) ^{2}%
\mathcal{\varpi }_{2}^{2}+\left( \sin \varphi \sin \phi -\cos \varphi \cos
\phi \sin \theta \right) ^{2}\mathcal{\varpi }_{3}^{2}  \notag \\
&&+2\cos \theta \cos \varphi \left[ \left( \cos \phi \sin \varphi +\sin
\theta \sin \phi \cos \varphi \right) K_{12}+\left( \sin \varphi \sin \phi
-\cos \varphi \cos \phi \sin \theta \right) K_{13}\right]  \notag \\
&&+2\left[ \cos \phi \sin \phi \left( \sin ^{2}\varphi -\cos ^{2}\varphi
\sin ^{2}\theta \right) -\sin \theta \sin \phi \cos \varphi \cos \left(
2\phi \right) \right] K_{23}
\end{eqnarray}

\begin{eqnarray}
M_{12} &=&-\cos ^{2}\theta \cos \varphi \sin \varphi \mathcal{\varpi }%
_{1}^{2}+\left[ \sin \varphi \cos \varphi \left( \cos ^{2}\phi -\sin
^{2}\theta \sin ^{2}\phi \right) +\cos \phi \sin \phi \sin \theta \cos
\left( 2\varphi \right) \right] \mathcal{\varpi }_{2}^{2}  \notag \\
&&+\left[ \cos \varphi \sin \varphi \left( \sin ^{2}\phi -\sin ^{2}\theta
\cos ^{2}\phi \right) -\cos \phi \sin \phi \sin \theta \cos \left( 2\varphi
\right) \right] \mathcal{\varpi }_{3}^{2}  \notag \\
&&+\cos \theta \left[ \cos \phi \cos \left( 2\varphi \right) -2\cos \varphi
\sin \phi \sin \varphi \sin \theta \right] K_{12}  \notag \\
&&+\cos \theta \left[ \sin \phi \cos \left( 2\varphi \right) +2\cos \varphi
\cos \phi \sin \theta \sin \varphi \right] K_{13}  \notag \\
&&+\left[ 2\left( \sin ^{2}\theta +1\right) \cos \varphi \cos \phi \sin
\varphi \sin \phi -\sin \theta \cos \left( 2\phi \right) \cos \left(
2\varphi \right) \right] K_{23}
\end{eqnarray}%
\begin{eqnarray}
M_{13} &=&\sin \theta \cos \theta \cos \varphi \mathcal{\varpi }%
_{1}^{2}-\sin \phi \cos \theta \left( \cos \phi \sin \varphi +\sin \theta
\sin \phi \cos \varphi \right) \mathcal{\varpi }_{2}^{2}+\cos \phi \cos
\theta (\sin \varphi \sin \phi -  \notag \\
&&\cos \varphi \cos \phi \sin \theta )\mathcal{\varpi }_{3}^{2}+\left[ \cos
\phi \sin \varphi \sin \theta -\sin \phi \cos \varphi \cos \left( 2\theta
\right) \right] K_{12}  \notag \\
&&+\left[ \sin \phi \sin \varphi \sin \theta +\cos \phi \cos \varphi \cos
\left( 2\theta \right) \right] K_{13}  \notag \\
&&+\left[ 2\sin \theta \sin \phi \cos \varphi \cos \phi \cos \theta +\sin
\varphi \cos \theta \cos \left( 2\phi \right) \right] K_{23}\text{ }
\end{eqnarray}%
\begin{eqnarray}
M_{21} &=&-\cos ^{2}\theta \sin \varphi \cos \varphi \mathcal{\varpi }%
_{1}^{2}+\left[ \cos \phi \sin \phi \sin \theta \cos \left( 2\varphi \right)
+\sin \varphi \cos \varphi \left( \cos ^{2}\phi -\sin ^{2}\phi \sin
^{2}\theta \right) \right] \mathcal{\varpi }_{2}^{2}  \notag \\
&&+\left[ -\cos \phi \sin \phi \sin \theta \cos \left( 2\varphi \right)
+\sin \varphi \cos \varphi \left( \sin ^{2}\phi -\cos ^{2}\phi \sin
^{2}\theta \right) \right] \mathcal{\varpi }_{3}^{2}  \notag \\
&&+\cos \theta \left[ \cos \phi \cos \left( 2\varphi \right) -2\cos \varphi
\sin \phi \sin \varphi \sin \theta \right] K_{12}  \notag \\
&&+\cos \theta \left[ \sin \phi \cos \left( 2\varphi \right) +2\cos \varphi
\sin \theta \cos \phi \sin \varphi \right] K_{13}  \notag \\
&&+\left[ \sin \theta \left( \sin ^{2}\varphi \cos \left( 2\phi \right)
-\cos ^{2}\varphi \cos \left( 2\phi \right) \right) +2\left( 1+\sin
^{2}\theta \right) \cos \varphi \sin \varphi \cos \phi \sin \phi \right]
K_{23}
\end{eqnarray}%
\begin{eqnarray}
M_{22} &=&\cos ^{2}\theta \sin ^{2}\varphi \mathcal{\varpi }_{1}^{2}+\left(
\cos \phi \cos \varphi -\sin \phi \sin \varphi \sin \theta \right) ^{2}%
\mathcal{\varpi }_{2}^{2}+\left( \cos \varphi \sin \phi +\sin \theta \cos
\phi \sin \varphi \right) ^{2}\mathcal{\varpi }_{3}^{2}  \notag \\
&&-2\cos \theta \sin \varphi \left( \cos \phi \cos \varphi -\sin \phi \sin
\varphi \sin \theta \right) K_{12}-2\cos \theta \sin \varphi \left( \cos
\varphi \sin \phi +\sin \theta \cos \phi \sin \varphi \right) K_{13}  \notag
\\
&&+2\left[ \sin \phi \cos \phi \left( \cos ^{2}\varphi -\sin ^{2}\varphi
\sin ^{2}\theta \right) +\sin \theta \cos \varphi \sin \varphi \cos \left(
2\phi \right) \right] K_{23}
\end{eqnarray}%
\begin{eqnarray}
M_{23} &=&-\sin \theta \cos \theta \sin \varphi \mathcal{\varpi }%
_{1}^{2}+\sin \phi \cos \theta \left( \sin \phi \sin \varphi \sin \theta
-\cos \phi \cos \varphi \right) \mathcal{\varpi }_{2}^{2}+\cos \phi \cos
\theta (\cos \varphi \sin \phi +  \notag \\
&&\sin \theta \cos \phi \sin \varphi )\mathcal{\varpi }_{3}^{2}+\left[ \cos
\phi \cos \varphi \sin \theta +\sin \varphi \sin \phi \cos \left( 2\theta
\right) \right] K_{12}+  \notag \\
&&\left[ \cos \varphi \sin \phi \sin \theta -\sin \varphi \cos \phi \cos
\left( 2\theta \right) \right] K_{13}+\cos \theta \left[ \cos \varphi \cos
\left( 2\phi \right) -2\sin \phi \sin \theta \cos \phi \sin \varphi \right]
K_{23}\text{ \ \ \ \ \ \ }
\end{eqnarray}%
\begin{eqnarray}
M_{31} &=&\cos \theta \sin \theta \cos \varphi \mathcal{\varpi }%
_{1}^{2}-\cos \theta \sin \phi \left( \sin \phi \sin \theta \cos \varphi
+\cos \phi \sin \varphi \right) \mathcal{\varpi }_{2}^{2}  \notag \\
&&+\cos \theta \cos \phi \left( \sin \varphi \sin \phi -\cos \phi \sin
\theta \cos \varphi \right) \mathcal{\varpi }_{3}^{2}  \notag \\
&&+\left[ \sin \theta \cos \phi \sin \varphi -\sin \phi \cos \varphi \cos
\left( 2\theta \right) \right] K_{12}+\left[ \cos \varphi \cos \phi \cos
\left( 2\theta \right) +\sin \theta \sin \varphi \sin \phi \right] K_{13} 
\notag \\
&&+\cos \theta \left[ 2\cos \phi \sin \phi \sin \theta \cos \varphi +\sin
\varphi \cos \left( 2\phi \right) \right] K_{23}
\end{eqnarray}%
\begin{eqnarray}
M_{32} &=&-\cos \theta \sin \theta \sin \varphi \mathcal{\varpi }%
_{1}^{2}-\cos \theta \sin \phi \left( \cos \phi \cos \varphi -\sin \phi \sin
\varphi \sin \theta \right) \mathcal{\varpi }_{2}^{2}  \notag \\
&&+\cos \phi \cos \theta \left( \sin \phi \cos \varphi +\sin \varphi \cos
\phi \sin \theta \right) \mathcal{\varpi }_{3}^{2}  \notag \\
&&+\left[ \sin \phi \sin \varphi \cos \left( 2\theta \right) +\sin \theta
\cos \phi \cos \varphi \right] K_{12}  \notag \\
&&+\left[ -\cos \phi \sin \varphi \cos \left( 2\theta \right) +\sin \theta
\sin \phi \cos \varphi \right] K_{13}  \notag \\
&&+\cos \theta \left[ \cos \varphi \cos \left( 2\phi \right) -2\cos \phi
\sin \phi \sin \varphi \sin \theta \right] K_{23}
\end{eqnarray}%
\begin{eqnarray}
M_{33} &=&\sin ^{2}\theta \mathcal{\varpi }_{1}^{2}+\cos ^{2}\theta \sin
^{2}\phi \mathcal{\varpi }_{2}^{2}+\cos ^{2}\phi \cos ^{2}\theta \mathcal{%
\varpi }_{3}^{2}  \notag \\
&&-2\cos \theta \left[ \sin \theta \sin \phi K_{12}-\sin \theta \cos \phi
K_{13}+\cos \theta \cos \phi \sin \phi K_{23}\right]
\end{eqnarray}

\end{document}